\documentclass[intlimits,twoside,a4paper]{article}

\usepackage{amsmath,amssymb}
\usepackage{graphicx}
\usepackage[T2A]{fontenc}
\usepackage[cp1251]{inputenc}

\usepackage{graphicx}

\usepackage[eqsecnum]{cmpj2}

\issue{2013}{16}{2}{23002}

\doinumber{10.5488/CMP.16.23002}

\newcommand{\be}{\begin{eqnarray}}
\newcommand{\ee}{\end{eqnarray}}
\newcommand{\Dlt}{\Delta}
\newcommand{\dlt}{\delta}
\newcommand{\prt}{\partial}
\newcommand{\br}{{\bf r}}

\newcommand{\bk}{{\bf k}}

\newcommand{\bp}{{\bf p}}
\newcommand{\ba}{{\bf a}}

\newcommand{\bP}{{\bf P}}
\newcommand{\bt}{\beta}
\newcommand{\vp}{\varphi}
\newcommand{\ep}{\varepsilon}
\newcommand{\al}{\alpha}
\newcommand{\ra}{\rightarrow}
\newcommand{\sgm}{\sigma}

\newcommand{\om}{\omega}
\newcommand{\Om}{\Omega}
\newcommand{\Gm}{\Gamma}
\newcommand{\dgr}{\dagger}
\newcommand{\lbd}{\lambda}
\newcommand{\Lbd}{\Lambda}

\newcommand{\cB}{{\cal B}}

\newcommand{\lgl}{\langle}
\newcommand{\rgl}{\rangle}

\title[Bose-condensed atoms in optical lattices]%
{Self-consistent approach for Bose-condensed atoms in optical lattices}
\author[V.I. Yukalov]{V.I. Yukalov}
\address{Bogolubov Laboratory of Theoretical Physics,
Joint Institute for Nuclear Research, 141980 Dubna, Russia
}

\date{Received December 7, 2012, in final form March 6, 2013}
\authorcopyright{V.I. Yukalov, 2013}

\begin{document}

\maketitle

\begin{abstract}
Bose atoms in optical lattices are considered at low temperatures
and weak interactions, when Bose-Einstein condensate is formed. A
self-consistent approach, based on the use of a representative
statistical ensemble, is employed, ensuring a gapless spectrum
of collective excitations and the validity of conservation laws. In
order to show that the approach is applicable to both weak and tight
binding, the problem is treated in the Bloch as well as in the
Wannier representations. Both these ways result in similar
expressions that are compared for the self-consistent
Hartree-Fock-Bogolubov approximation. A convenient general formula
for the superfluid fraction of atoms in an optical lattice is
derived.
\pacs 03.75.Hh, 03.75.Nt, 05.30.Ch, 05.30.Jp, 05.70.Ce
\end{abstract}

\section{Introduction}

Systems with Bose-Einstein condensate are interesting objects from both
theoretical and experimental points of view. That is why they
have been intensively studied in recent years. Vast literature on this problem
can be found in the books \cite{Pitaevskii_1,Lieb_2,Letokhov_3,Pethick_4} and
review articles \cite{Courteille_5,Andersen_6,Yukalov_7,Bongs_8,Yukalov_9,
Posazhennikova_10,Yukalov_11,Proukakis_12,Yurovsky_13,Yukalov_14,Yukalov_15}.
Creation of optical lattices has made it possible to achieve a new dimension
in the physics of cold atoms, providing an opportunity for numerous novel
applications and for modeling many effects typical of condensed matter
\cite{Morsch_16,Moseley_17,Bloch_18,Yukalov_19}.

The occurrence of Bose-Einstein condensate is intimately related to the global
gauge symmetry breaking \cite{Lieb_2,Yukalov_11} that is a necessary and
sufficient condition for Bose-Einstein condensation. In the theory of
Bose-condensed systems, there exists an old problem, formulated by
Hohenberg and Martin \cite{Hohenberg_20}, who showed that, as soon as gauge
symmetry is broken, the description of such a system suffers from one of the
defects, either yielding unphysical spectrum of collective excitations or
resulting in broken conservation laws and incorrect thermodynamics. Any of these
deficiencies implies that the description is not self-consistent, corresponding
to an unstable system. This problem has been solved by employing representative
statistical ensembles \cite{Gibbs_21,Yukalov_22,Yukalov_23} to systems with a broken
gauge symmetry \cite{Yukalov_24,Yukalov_25,Yukalov_26,Yukalov_27}. This approach
was shown to be completely self-consistent and gapless, with the
Hartree-Fock-Bogolubov (HFB) approximation \cite{Bogolubov_28,Bogolubov_29}
providing an accurate description for uniform Bose systems \cite{Yukalov_27,Yukalov_30,
Yukalov_31,Yukalov_32}, as well as for these systems in random external potentials
\cite{Yukalov_33,Yukalov_34}.

In the present paper, this self-consistent approach is applied to Bose-condensed
atoms in optical lattices. Sections~2 and 3, contain the main definitions related
to optical lattices and Bose-condensed atoms, respectively. In section~4, the Bloch
representation is used, which can be more suitable for weak binding, while in
section~5, the Wannier representation is employed, which is more convenient for tight
binding. Both these cases are treated in the HFB approximation leading to similar
results. However, the Wannier representation, yielding the Hubbard Hamiltonian, is
a bit simpler. Some thermodynamic characteristics are considered in section~6, where
a general and convenient formula for superfluid fraction is derived. Section~7
concludes.

Throughout the paper, the system of units is used, where the Planck and Boltzmann
constants are set to be one.

\section{Optical lattices}

Optical lattices are created by laser beams forming standing waves, which
corresponds to the formation of a periodic lattice potential
\be
\label{2.1}
 V_\mathrm{L}(\br +\ba) = V_\mathrm{L}(\br) \,  ,
\ee
with ${\bf a}$ being a lattice vector with the components
$a_\alpha = \lambda_\alpha / 2$, where $\lambda_\alpha$ is a laser wavelength
and $\alpha = 1,2, \ldots, d$ enumerates spatial components in a $d$-dimensional
space. The standard form of the lattice potential is
\be
\label{2.2}
 V_\mathrm{L}(\br) = \sum_{\al=1}^d V_\al \sin^2(k_0^\al r_\al) \,  ,
\ee
with the laser wave vector
\be
\label{2.3}
 \bk_0 = \left \{ k_0^\al = \frac{2\pi}{\lbd_\al} = \frac{\pi}{a_\al}
\right \} \,  .
\ee
The lattice depth is defined by the parameter
\be
\label{2.4}
V_0 \equiv \frac{1}{d} \, \sum_{\al=1}^d V_\al \,  .
\ee
Another important quantity, characterizing an optical lattice, is the recoil energy
\be
\label{2.5}
 E_\mathrm{R} \equiv \frac{k_0^2}{2m} \,  , \qquad
k_0^2 \equiv \sum_{\al=1}^d (k_0^\al )^2 \, ,
\ee
where $m$ is atomic mass. The ratio $E_\mathrm{R}/V_0$ characterizes the relative lattice
depth.

\section{Bose atoms}

The lattice is loaded with Bose atoms, whose interactions are measured by means
of the scattering length $a_\mathrm{s}$ entering the effective interaction strength
\be
\label{3.1}
 \Phi_0 \equiv 4\pi \, \frac{a_\mathrm{s}}{m} \,  .
\ee
The energy operator is given by the  Hamiltonian
\begin{eqnarray}
\hat H = \int \hat\psi^\dgr(\br) \left ( -\, \frac{\nabla^2}{2m} +
U + V_\mathrm{L} \right ) \hat\psi(\br) \, \rd\br \,
+ \, \frac{1}{2} \Phi_0 \int  \hat\psi^\dgr(\br) \hat\psi^\dgr(\br)
\hat\psi(\br)  \hat\psi(\br) \, \rd\br \,   ,\label{3.2}
\end{eqnarray}
in which $U = U({\bf r})$ is a trapping potential, if any, and $V_\mathrm{L} = V_\mathrm{L}({\bf r})$
is a lattice potential. The atom field operators $\hat{\psi}({\bf r})$ satisfy the
Bose commutation relations.

The existence of Bose-Einstein condensate necessarily requires that global gauge
symmetry should be broken \cite{Lieb_2,Yukalov_11}. The most straightforward way of
the gauge symmetry breaking is by means of the Bogolubov shift of the field
operator
\be
\label{3.3}
 \hat\psi(\br) = \eta(\br) + \psi_1(\br) \,  .
\ee
Here, the first term is the condensate wave function normalized to the number of
condensed atoms
\be
\label{3.4}
 N_0 = \int | \eta(\br) |^2 d\br \,  .
\ee
The second term is the field operator of uncondensed atoms, whose number is
given by the statistical average
\be
\label{3.5}
N_1 = \lgl \hat N_1 \rgl \ , \qquad
\hat N_1 \equiv \int \psi_1^\dgr(\br) \psi_1(\br)\, \rd\br
\ee
of the number-of-particle operator $\hat{N}_1$.

The uncondensed atoms are normal in the sense that the average of their field
operator is zero,
\be
\label{3.6}
 \lgl \psi_1 \rgl = 0 \,  .
\ee
To avoid double counting of the degrees of freedom, the orthogonality condition
\be
\label{3.7}
\int \eta^*(\br) \psi_1(\br) \, \rd\br = 0
\ee
is required. This condition is a direct consequence of  orthogonality
of wave functions serving as a basis for the expansion of the field
operator $\hat\psi(\br)$ \cite{Bogolubov_28,Bogolubov_29}.

The number of atoms per lattice site is called a filling factor that is defined as
the ratio
\be
\label{3.8}
 \nu \equiv \frac{N}{N_\mathrm{L}} = \rho a^d \qquad ( N = N_0 + N_1 ) \,  ,
\ee
in which $a$ is a mean interatomic distance and $\rho$ is the average atomic density,
\be
\label{3.9}
a \equiv \left ( \frac{V}{N_\mathrm{L}} \right )^{1/d} \, , \qquad
\rho \equiv \frac{N}{V} \,   .
\ee

The representative ensemble for a system with a broken gauge symmetry is
characterized \cite{Yukalov_24,Yukalov_25,Yukalov_26,Yukalov_27} by the grand
Hamiltonian
\be
\label{3.10}
H = \hat H - \mu_0 N_0 - \mu_1 \hat N_1 - \hat\Lbd \,   ,
\ee
where $\mu_0$ and $\mu_1$ are the Lagrange multipliers ensuring the validity
of normalizations (3.4) and (3.5), while the term $\hat\Lambda$ is defined
so that the terms linear in the operators $\psi_1$ are cancelled in the Hamiltonian,
which ensures the condition (3.6).

It is worth stressing that the introduction of two Lagrange
multipliers, $\mu_0$ and $\mu_1$ is necessary due to the
presence of two independent variables in the Bogolubov shift (3.3)
and the related two normalization conditions (3.4) and (3.5). It is
a general mathematical fact that the number of Lagrange
multipliers should be equal to the number of imposed constraints,
such as the normalization conditions. The theory can become non-self-consistent if the number of Lagrange multipliers is smaller
than that of the imposed constraints. Introducing two Lagrange
multipliers does not exclude that in particular cases, these
multipliers could become equal, as it happens in the Bogolubov
approximation \cite{Bogolubov_28,Bogolubov_29}. The physical meaning
of using two Lagrange multipliers has been thoroughly explained in
the previous papers
\cite{Yukalov_11,Yukalov_14,Yukalov_19,Yukalov_23,Yukalov_24,Yukalov_25,
Yukalov_26,Yukalov_27,Yukalov_30,Yukalov_31,Yukalov_32,Yukalov_33}.

\section{Bloch representation}

One usually considers optical lattices by reducing the problem to a Hubbard
Hamiltonian by means of the Wannier representation which is convenient in the
case of a tight binding. Here, we show that it is equivalently possible to
employ the Bloch representation that can be more appropriate for weak binding
and leads to the results similar to those in the Wannier representation to be considered in the following section. Below, we assume that there is
no trapping potential, so that the system is ideally periodic.

Let $\{\varphi_{nk}({\bf r})\}$ be the basis of Bloch functions labeled by the
zone index $n$ and quasi-momentum multi-index $k$. Then, the field operators
of uncondensed atoms can be expanded over this basis,
\be
\label{4.1}
\psi_1(\br) = \sum_{nk} a_{nk} \vp_{nk}(\br) \,  .
\ee
The basis should be chosen so that the Bloch functions are natural orbitals
\cite{Coleman_35}, that is, the eigenfunctions of the
density matrix
\be
\label{4.2}
\rho_1(\br,\br')  \equiv \lgl \psi_1^\dgr(\br') \psi_1(\br) \rgl \,   .
\ee
Then, the density matrix enjoys a diagonal expansion
\be
\label{4.3}
 \rho_1(\br,\br') = \sum_{nk} \lgl a_{nk}^\dgr a_{nk} \rgl
\vp_{nk}(\br)\vp_{nk}^*(\br') \,  .
\ee
In other words, the use of natural orbitals simplifies the consideration
due to the following properties
\be
\label{4.4}
 \lgl a_{nk}^\dgr a_{mp} \rgl  =
\dlt_{mn}\dlt_{kp} \lgl a_{nk}^\dgr a_{nk} \rgl \, , \qquad
\lgl a_{nk} a_{mp} \rgl  =
\dlt_{mn}\dlt_{-kp} \lgl a_{nk} a_{np} \rgl \, .
\ee

Substituting expansion (4.1) into the grand Hamiltonian (3.10) gives the sum
\be
\label{4.5}
 H = H^{(0)} + H^{(2)} + H^{(3)} + H^{(4)} \,  .
\ee
Here, the first term
\be \label{4.6} H^{(0)} = \int \eta^*(\br)
\left ( - \, \frac{\nabla^2}{2m} + V_\mathrm{L} - \mu_0 \right ) \eta(\br) \,
\rd\br \, + \, \frac{1}{2} \, \Phi_0 \int |\eta(\br)|^4 \rd\br
\ee
contains only a condensate wave function, but no field operators
of uncondensed atoms.  The term, linear in $\psi_1$, is canceled by
the Lagrange term $\hat\Lambda$. In the following expressions, the
pair $\{n,k\}$, for brevity, will be denoted as $k$, while the set
$\{n,-k\}$, as $-k$. Then, the term, containing the products of two
operators of uncondensed atoms, reads as
\begin{eqnarray}
 H^{(2)} &=& \sum_{kp} \left [ \int \vp_k^*(\br) \left ( - \,
\frac{\nabla^2}{2m} + V_\mathrm{L} - \mu_1 + 2\Phi_0 |\eta(\br)|^2
\right ) \vp_p(\br) \, \rd\br \right ] a_k^\dgr a_p  \, \nonumber\\
&&
+ \, \frac{1}{2} \sum_{kp} \left ( \Phi_{kp}a_k^\dgr a_p^\dgr +
\Phi_{kp}^* a_p a_k \right )   \, ,
\label{4.7}
\end{eqnarray}
where
$$
\Phi_{kp} \equiv
\Phi_0 \int \vp_k^*(\br) \vp_p^*(\br) \eta^2(\br) \, \rd\br \,   .
$$
The term of third order, with respect to the products of the field operators
of uncondensed atoms, is
\be
\label{4.8}
 H^{(3)} = \sum_{kpq} \left ( \int \Phi_{kpq} a_k^\dgr a_p^\dgr a_q
+  \Phi^*_{kpq} a_q^\dgr a_p a_k \right )\, ,
\ee
with
$$
\Phi_{kpq} \equiv
\Phi_0 \int \vp_k^*(\br) \vp_p^*(\br)\vp_q(\br)\eta(\br) \, \rd\br \, .
$$
And the fourth-order term is
\be
\label{4.9}
 H^{(4)} = \frac{1}{2} \, \sum_{kpql}
\Phi_{kpql} a_k^\dgr a_p^\dgr a_q a_l \, ,
\ee
where
$$
\Phi_{kpql} \equiv
\Phi_0 \int \vp_k^*(\br) \vp_p^*(\br)\vp_q(\br)\vp_l(\br) \, \rd\br \, .
$$

In the Hartree-Fock-Bogolubov (HFB) approximation, the third-order
term $H^{(3)}$ yields expressions linear in $\psi_1$, which should
be canceled by the Lagrange canceler $\hat\Lambda$. The fourth-order
part takes the form
\begin{eqnarray}
H^{(4)} &=& \frac{1}{2} \sum_{kpq} \left ( 4 \Phi_{kqqp} n_q a_k^\dgr a_p +
\Phi_{kpqq} \sgm_q a_k^\dgr a_p^\dgr + \Phi_{kpqq}^* \sgm_q^* a_p a_k \right )\nonumber\\
&& - \, \frac{1}{2} \sum_{kp} \left ( 2 \Phi_{kppk} n_k n_p +
\Phi_{kkpp}\sgm^*_k \sgm_p \right ) \,   ,
\label{4.10}
\end{eqnarray}
in which the notations for the so-called normal
\be
\label{4.11}
 n_k \equiv \lgl a_k^\dgr a_k \rgl \,   ,
\ee
and anomalous
\be
\label{4.12}
 \sgm_k \equiv \lgl a_k a_{-k} \rgl \,
\ee
averages are used. The normal average (4.11) is the distribution of atoms,
while the absolute value $|\sigma_k|$ of the anomalous average (4.12) is the
distribution of the correlated atomic pairs
\cite{Yukalov_19,Yukalov_25,Bogolubov_29}.

Let us introduce the notation
\be
\label{4.13}
\om_{kp} \equiv \int \vp_k^*(\br) \left ( - \, \frac{\nabla^2}{2m} + V_\mathrm{L} +
2\Phi_0 |\eta|^2 \right ) \vp_p(\br) \, \rd\br \, +
\, 2 \sum_q \Phi_{kqqp} n_q \, - \, \mu_1 \dlt_{kp}
\ee
and
\be
\label{4.14}
\Dlt_{kp} \equiv \Phi_{kp} + \sum_q \Phi_{kpqq} \sgm_q \, .
\ee
Then, the grand Hamiltonian (4.5) in the HFB approximation can be written as
\be
\label{4.15}
H = E_\mathrm{HFB} + \sum_{kp} \om_{kp} a_k^\dgr a_p \, + \,
\frac{1}{2} \, \sum_{kp} \left ( \Dlt_{kp} a_k^\dgr a_p^\dgr
+\Dlt^*_{kp} a_p a_k \right ) \,  ,
\ee
where the first term is the nonoperator quantity
\be
\label{4.16}
 E_\mathrm{HFB} = H^{(0)} \, - \, \frac{1}{2} \sum_{kp} ( 2 \Phi_{kppk} n_k n_p
+ \Phi_{kkpp} \sgm_k^* \sgm_p ) \,  .
\ee

The quadratic Hamiltonian (4.15) can be diagonalized and all observables
calculated. However, the resulting expressions are rather complicated. In order
to simplify the calculations, it is possible to assume that the main contribution
in the above formulas comes from diagonal terms, since the Bloch functions are
mutually orthogonal. This can be referred to as the {\it diagonal approximation}, when
expressions (4.13) and (4.14) take the form
\be
\label{4.17}
 \om_{kp} = \dlt_{kp} \om_k \, , \qquad
\Dlt_{kp} = \dlt_{-kp} \Dlt_k \,  ,
\ee
in which
\be
\label{4.18}
\om_k = \int \vp_k^*(\br) \left ( -\, \frac{\nabla^2}{2m} + V_\mathrm{L} +
2\Phi_0 |\eta(\br)|^2 \right ) \vp_k(\br)\, \rd\br \, + \,
2 \sum_q \Phi_{kqqk} n_q \, - \, \mu_1
\ee
and
\be
\label{4.19}
\Dlt_k = \Phi_{-kk} + \sum_q \Phi_{-kkqq}\sgm_q \,   .
\ee

The use of the diagonal approximation is not compulsory and it is possible
to diagonalize the quadratic form (4.15) without it. This approximation,
however, essentially simplifies the formulas. Justification of this
approximation is based on the fact that the expansion functions $\vp_k$
are mutually orthogonal, which makes it reasonable to assume that the
matrix elements over these functions are such that their diagonal elements
are larger than off-diagonal.

In the diagonal approximation, Hamiltonian (4.15) reduces to
\be
\label{4.20}
 H = E_\mathrm{HFB} + \sum_k \om_k a_k^\dgr a_k \, + \, \frac{1}{2}
\sum_k \left ( \Dlt_k a_k^\dgr a_{-k}^\dgr +
\Dlt_k^* a_{-k} a_k \right ) \,  .
\ee
This form is much simpler to diagonalize using the Bogolubov canonical
transformation \cite{Bogolubov_28,Bogolubov_29}.

Following a standard procedure by diagonalizing Hamiltonian (4.20), we
find the Bogolubov spectrum of elementary excitations
\be
\label{4.21}
 \ep_k =\sqrt{\om_k^2 -\Dlt_k^2} \,  .
\ee
The condition of condensate existence \cite{Yukalov_14,Yukalov_19} requires
that the spectrum should be gapless,
\be
\label{4.22}
\lim_{k\ra 0}\ep_k = 0 \, , \qquad \ep_k \geqslant 0 \,   .
\ee
This condition is equivalent to the Hugenholtz-Pines theorem \cite{Hugenholtz_36}.
Hence, we get
\begin{eqnarray}
\mu_1 &=& \frac{1}{N_0} \int \eta^*(\br) \left\{ -\,
\frac{\nabla^2}{2m} + V_\mathrm{L}(\br) + \Phi_0 [ \rho_0(\br) + 2\rho_1(\br) ]
\right \} \eta(\br)\, \rd\br \, \nonumber\\
&&-
\frac{\Phi_0}{N_0} \int \sgm_1(\br) [\eta^*(\br)]^2 \rd\br \,  ,
\label{4.23}
\end{eqnarray}
where the notations are used for the condensate density
\be
\label{4.24}
 \rho_0(\br) \equiv |\eta(\br)|^2 \,   ,
\ee
density of uncondensed atoms
\be
\label{4.25}
 \rho_1(\br) \equiv \sum_k n_k |\vp_k(\br) |^2 \, ,
\ee
and the anomalous average
\be
\label{4.26}
 \sgm_1(\br) \equiv \sum_k \sgm_k\vp_k(\br) \vp_{-k}(\br) \,  .
\ee

The equation for the condensate wave function, in the case of an equilibrium
system, is defined by the variational condition
\be
\label{4.27}
 \left\lgl \frac{\dlt H}{\dlt\eta^*(\br)} \right \rgl = 0 \,  ,
\ee
which yields the equation
\be
\label{4.28}
 \left \{ - \, \frac{\nabla^2}{2m} + V_\mathrm{L}(\br) +
\Phi_0 [ \rho_0(\br) + 2\rho_1(\br) ]\right \} \eta(\br) +
\Phi_0\sgm_1(\br)\eta^*(\br) = \mu_0\eta(\br) \, .
\ee
The latter gives the condensate chemical potential
\be
\label{4.29}
\mu_0 &=& \frac{1}{N_0} \int \eta^*(\br) \left\{ -\,
\frac{\nabla^2}{2m} + V_\mathrm{L}(\br) + \Phi_0 [ \rho_0(\br) + 2\rho_1(\br) ]
\right \} \eta(\br)\, \rd\br \, \nonumber\\
&& + \frac{\Phi_0}{N_0} \int \sgm_1(\br) [\eta^*(\br)]^2 \rd\br \, .
\ee
Comparing expressions (\ref{4.23}) and ({4.29}), we see that they are connected by
the relation
\be
\label{4.30}
 \mu_0 = \mu_1 +
\frac{2\Phi_0}{N_0} \int \sgm_1(\br) [\eta^*(\br)]^2 \rd\br \,  .
\ee

Evidently, the Lagrange multipliers $\mu_0$ and $\mu_1$ do not coincide.
The system chemical potential is defined through the equation
\be
\label{4.31}
 \lgl H \rgl = \lgl \hat H \rgl - \mu N \,  ,
\ee
which yields
\be
\label{4.32}
 \mu = \frac{1}{N} ( \lgl \hat H \rgl - \lgl H \rgl ) \,  .
\ee
This leads to the expression
\be
\label{4.33}
  \mu = \mu_0 n_0 + \mu_1 n_1 \, ,
\ee
in which the condensate fraction $n_0$ and the fraction of uncondensed atoms,
$n_1$, are introduced,
$$
n_0 \equiv \frac{N_0}{N} \, , \qquad n_1 \equiv \frac{N_1}{N} \, .
$$
Invoking equation~(\ref{4.30}), we get
\be
\label{4.34}
 \mu = \mu_1 + \frac{2\Phi_0}{N}
\int \sgm_1(\br) [ \eta^*(\br)]^2 \, \rd\br \,  .
\ee

Sometimes, one requires that $\mu$ should be equal to $\mu_0$ and $\mu_1$, which forces us
to assume that the anomalous average $\sigma_1$ should be zero. Such a requirement has
no physical reason. In addition, it can be shown by direct calculations
\cite{Yukalov_14,Yukalov_19,Yukalov_37} that the anomalous average is always
comparable with or larger than either the density of uncondensed atoms or
that of condensed atoms. Therefore, there is no such a region of parameters,
where it could be admissible to neglect the anomalous average, but to keep the
normal density and the density of condensed atoms. The sole possibility could
be at temperatures close to zero and asymptotically weak interactions, when,
though the anomalous average is three times larger than the normal density,
both of them are much smaller than the condensate density. Then, it could be
possible to omit both the anomalous average and the normal density,
keeping only the condensate density. But neglecting one of them, though keeping
another one, is mathematically wrong. Moreover, neglecting the anomalous average
is not merely mathematically incorrect, but it is qualitatively deficient,
making thermodynamics non-self-consistent, disturbing the condensate transition
to the first order, and resulting in unphysical divergences of compressibility and
structure factor \cite{Yukalov_38}.

Since this section is based on the Bloch representation, it is necessary
to briefly describe how the Bloch functions could be defined. Formally, as has
been mentioned above, the basis of Bloch functions should be chosen as a
set of natural orbitals \cite{Coleman_35}, since this gives a diagonal expansion
for the density matrix (\ref{4.3}). However, the problem is that the density matrix (\ref{4.2})
is not known explicitly. Hence, it is impossible to find its exact eigenfunctions
representing the natural orbitals.  A standard way is to define the Bloch
functions as solutions to the equation
\be
\label{4.35}
 \left [ -\, \frac{\nabla^2}{2m} + V_\mathrm{L}(\br) \right ] \vp_{nk}(\br) =
E_{nk}\vp_{nk}(\br) \,  .
\ee

It is also possible to define Bloch functions as eigenfunctions of the
nonlinear Schr\"{o}dinger equation \cite{Yukalov_19}, including
the interaction term into equation~(\ref{4.35}). Then, calculations become essentially more complicated.
In addition, there arises a problem of nonorthogonality of
eigenfunctions of the nonlinear equation. Thus, the simplest way is to use the solutions to the linear equation~(\ref{4.35}) as a
basis, complimenting it by conservation conditions (\ref{4.4}).

\section{Wannier representation}

The field operator of atoms can be expanded over the basis of Wannier functions,
\be
\label{5.1}
 \hat\psi(\br) = \sum_{nj} \hat c_{nj} w_n(\br-\ba_j) \,   ,
\ee
where the index $n = 1,2, \ldots$ labels bands and $j = 1,2, \ldots, N_\mathrm{L}$
enumerates the lattice sites. Substituting this into Hamiltonian (\ref{3.2}),
considering just a single lowest band, and taking into account only the
nearest-neighbor interactions, one comes to the Hubbard model
\be
\label{5.2}
 \hat H = - J \sum_{\lgl ij\rgl} \hat c_i^\dgr \hat c_j \, + \,
\frac{U}{2} \sum_j \hat c_j^\dgr  \hat c_j^\dgr \hat c_j  \hat c_j \, + \,
h_0 \sum_j \hat c_j^\dgr \hat c_j \,  ,
\ee
here, the operators $\hat{c}_j$ satisfy the Bose commutation relations.

The parameters entering the Hubbard Hamiltonian (\ref{5.2}) can be calculated
in the tight-binding approximation. A detailed demonstration of this
calculation can be found in reference~\cite{Yukalov_19}. For a three-dimensional
space in this approximation, we find the expressions
\[
J = \frac{3}{4} \left ( \pi^2 - 4 \right ) V_0 \exp\left ( -\,
\frac{3\pi^2}{4}\,\sqrt{\frac{V_0}{E_\mathrm{R}} } \right ) \, , \qquad
U = \sqrt{\frac{8}{\pi} } \, k_0 a_\mathrm{s} E_\mathrm{R}\left (  \frac{V_0}{E_\mathrm{R}} \right )^{3/4},
\]
\vspace{-5mm}
\be
\label{5.3}
h_0 =  3 E_\mathrm{R} \, \sqrt{\frac{V_0}{E_\mathrm{R}} } \qquad (d = 3 )  \, .
\ee
The explanation of the notations for $V_0$, $E_\mathrm{R}$, and $k_0$ are given
in section~2.

The single-band Hamiltonian (\ref{5.2}) is called the boson Hubbard model. It is
possible to generalize this model by taking into account two or more bands
\cite{Yukalov_39,Stasyuk_40}. Here, we consider the single-band case, when the system displays Bose-Einstein condensation, though.

Employing the Bogolubov shift (\ref{3.3}), we have the condensate wave function
\be
\label{5.4}
 \eta(\br) = \sqrt{\nu n_0} \, \sum_j w(\br-\ba_j) \,  ,
\ee
with $n_0 = N_0/N$, and the operator of uncondensed atoms
\be
\label{5.5}
 \psi_1(\br) = \sum_j c_j w(\br-\ba_j) \,  .
\ee
In terms of the operators $c_j$, the Bogolubov shift reads as follows:
\be
\label{5.6}
 \hat c_j = \sqrt{\nu n_0} + c_j \,  .
\ee
Condition (\ref{3.6}) leads to the requirement
\be
\label{5.7}
 \lgl c_j \rgl = 0 \,  .
\ee
And from the orthogonality condition (\ref{3.7}), it follows that
\be
\label{5.8}
\sum_j c_j = 0 \,   .
\ee
The grand Hamiltonian (\ref{3.10}), with
$$
\hat \Lbd = \sum_j \left ( \lbd_j c_j^\dgr + \lbd_j^* c_j\right ) \, ,
$$
takes the form (\ref{4.5}). The constant $h_0$ can be incorporated into the chemical
potentials $\mu_0$ and $\mu_1$. The zero-order term is
\be
\label{5.9}
H^{(0)} = - J z_0 n_0 N + \frac{U}{2} \, \nu n_0^2 N - \mu_0 n_0 N \,   ,
\ee
where the number of the nearest neighbors is denoted as
\be
\label{5.10}
 z_0 \equiv \frac{1}{N} \, \sum_{\lgl ij\rgl } 1 \,  .
\ee The first-order term is canceled by the linear canceler
$\hat\Lambda$. The second-order term is \be \label{5.11}
 H^{(2)} = - J \sum_{\lgl ij\rgl } c_i^\dgr c_j +
(2 U \nu n_0 - \mu_1 ) \sum_j c_j^\dgr c_j +
\frac{U}{2} \, \nu n_0 \sum_j \left (  c_j^\dgr c_j^\dgr + c_j c_j \right ) \, .
\ee
The third-order term reads as follows:
\be
\label{5.12}
 H^{(3)} = U \sqrt{\nu n_0} \sum_j \left ( c_j^\dgr c_j^\dgr c_j +
 c_j^\dgr c_j c_j \right ) \, .
\ee
The fourth-order terms is
\be
\label{5.13}
 H^{(4)} = \frac{U}{2} \, \sum_j c_j^\dgr c_j^\dgr c_j c_j \, .
\ee

The fraction of uncondensed atoms takes the form
\be
\label{5.14}
 n_1 = \frac{1}{N} \, \sum_j \lgl c_j^\dgr c_j \rgl =
\frac{1}{\nu} \, \lgl c_j^\dgr c_j \rgl \,  ,
\ee
where the lattice ideality is used. For the dimensionless anomalous average,
we have
\be
\label{5.15}
\sgm = \frac{1}{N} \, \sum_j \lgl c_j c_j \rgl =
\frac{1}{\nu} \, \lgl c_j c_j \rgl \,    .
\ee

The necessary condition of the system stability
\be
\label{5.16}
\left \lgl \frac{\prt H}{\prt N_0} \right \rgl = 0
\ee
yields
\be
\label{5.17}
\mu_0 = - Jz_0 + \nu U \left [ n_0 + 2n_1 +
\frac{1}{2} \, (\sgm^* + \sgm ) \right ]  +
\frac{U}{2\sqrt{\nu n_0} } \sum_j \lgl c_j^\dgr c_j^\dgr c_j
+ c_j^\dgr c_j  c_j \rgl  \, .
\ee

The operators $c_j$ can be expanded over the Fourier basis,
\be
\label{5.18}
c_j = \frac{1}{\sqrt{N_\mathrm{L}} } \sum_k a_k \re^{\ri\bk\cdot\ba_j} \, ,
\ee
where $k$ runs over the Brillouin zone.

Let us consider a cubic lattice. Then, the second-order term (5.11) becomes
\be
\label{5.19}
 H^{(2)} = \sum_k \left [ - 2J \sum_{\al=1}^d \cos(k_\al a) +
2U \nu n_0 - \mu_1 \right ] a_k^\dgr a_k \, +  \,
\frac{U}{2}\, \nu n_0 \sum_k \left ( a_k^\dgr a_{-k}^\dgr +
a_{-k} a_k \right ) \, .
\ee
The third-order and fourth-order terms are
\be
\label{5.20}
H^{(3)} = U\, \sqrt{\frac{\nu n_0}{N_\mathrm{L}} } \,
\sum_{kp} \left ( a_k^\dgr a_p^\dgr a_{k+p} +
a_{k+p}^\dgr a_p a_k \right )
\ee
and, respectively,
\be
\label{5.21}
 H^{(4)} = \frac{U}{2N_\mathrm{L}} \,
\sum_{kpq} a_k^\dgr a_p^\dgr a_{k+p} a_{p-q} \,  .
\ee

In the HFB approximation, the third-order term is zero, due to condition
(\ref{5.8}). And the fourth-order term in the HFB approximation reads as follows:
\be
\label{5.22}
 H^{(4)} = \frac{\nu}{2} \, U \sum_k \left ( 4n_1 a_k^\dgr a_k +
\sgm a_k^\dgr a_{-k}^\dgr + \sgm^* a_{-k} a_k \right ) \, - \,
\frac{\nu}{2} \, U N \left ( 2n_1^2 + | \sgm|^2 \right ) \,  .
\ee
Introducing the notations
\be
\label{5.23}
\om_k \equiv -2J \sum_{\al=1}^d \cos(k_\al a) + 2\nu U - \mu_1
\ee
and
\be
\label{5.24}
 \Dlt \equiv \nu U (n_0 + \sgm) \,
\ee
for the grand Hamiltonian (\ref{4.5}), we obtain
\be
\label{5.25}
 H = E_\mathrm{HFB} + \sum_k \om_k a_k^\dgr a_k \, + \,
\frac{1}{2} \sum_k \left ( \Dlt a_k^\dgr a_{-k}^\dgr +
\Dlt^* a_{-k} a_k \right ) \,  ,
\ee
where
$$
E_\mathrm{HFB} \equiv H^{(0)} - \nu N \, \frac{U}{2} \left ( 2n_1^2 +
|\sgm|^2 \right ) \,   .
$$
The condensate chemical potential (\ref{5.17}) in the HFB approximation becomes
\be
\label{5.26}
 \mu_0 = -z_0 J + \nu U ( 1 + n_1 +\sgm) \,  .
\ee

Diagonalizing Hamiltonian (\ref{5.25}), we get the Bogolubov Hamiltonian
\be
\label{5.27}
 H_\mathrm{B} = E_\mathrm{B} + \sum_k \ep_k b_k^\dgr b_k \,  ,
\ee
in which
$$
E_\mathrm{B} = E_\mathrm{HFB} + \frac{1}{2} \sum_k (\ep_k - \om_k )\, ,
$$
and the Bogolubov spectrum is
\be
\label{5.28}
  \ep_k = \sqrt{\om_k^2-\Dlt^2 } \, .
\ee
The condition of the condensate existence (\ref{4.22}) yields
\be
\label{5.29}
\mu_1 = -z_0 J + \nu U ( 1 + n_1 - \sgm) \,   .
\ee
Then, equation (\ref{5.23}) becomes
\be
\label{5.30}
 \om_k = \Dlt +
4J \sum_{\al=1}^d \sin^2\left ( \frac{k_\al a}{2} \right ) \,  .
\ee
And, introducing the notation
\be
\label{5.31}
 e_k = 4J \sum_{\al=1}^d \sin^2\left ( \frac{k_\al a}{2} \right ) \,
\ee
for the Bogolubov spectrum (\ref{5.28}), we get
\be
\label{5.32}
  \ep_k = \sqrt{e_k(e_k+2\Dlt) } \,  .
\ee

Comparing equations (\ref{5.26}) and (\ref{5.29}) yields the relation
\be
\label{5.33}
\mu_0 = \mu_1 + 2\nu U \sgm \, .
\ee
As is seen, $\mu_0$ does not coincide with $\mu_1$, by analogy with relation
(\ref{4.30}). The anomalous average cannot be neglected, as is explained in section~4.

For the quasi-momentum atomic distribution and for the quasi-momentum
representation of the anomalous average, respectively, we find
\be
\label{5.34}
 n_k \equiv \lgl a_k^\dgr a_k \rgl =
\frac{\om_k}{2\ep_k} \, \coth\left ( \frac{\ep_k}{2T}\right ) -
\, \frac{1}{2} \, , \qquad
\sgm_k \equiv \lgl a_k a_{-k} \rgl = -\,
\frac{\Dlt}{2\ep_k} \, \coth\left ( \frac{\ep_k}{2T}\right ) \, .
\ee
This shows that the normal and anomalous averages are connected by the relation
\[
 \sgm_k^2 = n_k ( 1 + n_k )  -\, \frac{1}{4\sinh^2(\ep_k/2T)} \,  .
\]
For the integral quantities (\ref{5.14}) and (\ref{5.15}), we have
\[
 n_1 = \frac{1}{\rho} \int_{\cB} n_k \, \frac{\rd\bk}{(2\pi)^d} \, ,
\qquad
\sgm = \frac{1}{\rho} \int_{\cB} \sgm_k \, \frac{\rd\bk}{(2\pi)^d} \, .
\]
The condensate fraction reads as
\be
\label{5.35}
 n_0 =  1 - \, \frac{1}{2\rho} \int_{\cB} \left [
\frac{\om_k}{\ep_k} \, \coth\left ( \frac{\ep_k}{2T}\right ) - 1 \right ] \,
\frac{\rd\bk}{(2\pi)^d} \, ,
\ee
with the integration over the Brillouin zone.

Let us emphasize again that the anomalous average cannot be neglected
for the principal reason. As is evident form the above formulas, the anomalous
average can be zero only when there is no condensate, $n_0 = 0$. Hence, there
is no gauge symmetry breaking. However, as soon as there appears Bose-Einstein
condensate, the gauge symmetry becomes broken, and the anomalous average is never
zero. It is always comparable with or larger than either the density of
uncondensed atoms or that of condensed atoms.

\section{Thermodynamic characteristics}

In the HFB approximation, the grand potential takes the form
\be
\label{6.1}
\Om = E_\mathrm{B} + TV  \int_{\cB} \ln \left ( 1 - \re^{-\bt\ep_k}
\right ) \frac{\rd\bk}{(2\pi)^d} \,   ,
\ee
where the integration is over the Brillouin zone and
\[
 E_\mathrm{B} = H^{(0)} -\, \frac{N}{2} \,\nu U \left ( 2n_1^2 + \sgm^2 \right ) +
\frac{N}{2\rho} \int_{\cB} (\ep_k - \om_k ) \frac{\rd\bk}{(2\pi)^d} \, .
\]
The system chemical potential (\ref{4.33}) is
\be
\label{6.2}
 \mu = \mu_0 n_0 + \mu_1 n_1 =
- z_0 J + \nu U ( 1 + n_1 + \sgm - 2n_1\sgm) \,  .
\ee
For the ground-state energy
\be
\label{6.3}
 E_0 \equiv E_\mathrm{B} + \mu N \,  ,
\ee
we have
\be
\label{6.4}
  \frac{E_0}{N} = - z_0 J +
\frac{1}{2} \, \nu U \left ( 1 + n_1^2 -\sgm^2 - 2n_1\sgm \right ) +
\frac{1}{2\rho} \int_{\cB} (\ep_k - \om_k ) \frac{\rd\bk}{(2\pi)^d} \, .
\ee

Atomic fluctuations are characterized by the number-of-atom operator variance
\be
\label{6.5}
\mathrm{var}(\hat N) \equiv \lgl \hat N^2 \rgl -
\lgl \hat N \rgl^2 \,  ,
\ee
in which
\be
\label{6.6}
\hat N = N_0 + \hat N_1
\ee
is the operator of the total number of atoms. Since the first term $N_0$ is
a non-operator number, one has
\be
\label{6.7}
\mathrm{var}(\hat N) = {\rm var}(\hat N_1) .
\ee
In the HFB approximation, we get
\be
\label{6.8}
\mathrm{var}(\hat N_1)  = \frac{NT}{\nu U(n_0+\sgm)} \, .
\ee

The number-of-atom operator variance defines the isothermic compressibility
\be
\label{6.9}
 \kappa_T = \frac{{\rm var}(\hat N)}{\rho TN} =
\frac{1}{\rho\nu U(n_0+\sgm)} \,  .
\ee

The atomic fluctuations are, of course, normal and the compressibility is finite
everywhere below $T_\mathrm{c}$. The compressibility can diverge only at the critical
point $T_\mathrm{c}$.

Bose-Einstein condensation is a second-order phase transition occurring at a
temperature $T_\mathrm{c}$, where $n_0 = 0$ and $\sigma = 0$. At this point, the atomic
density is
\be
\label{6.10}
 \rho = \frac{1}{2} \int_{\cB} \left [
\coth \left ( \frac{\om_k}{2T_\mathrm{c}} \right ) -1
\right ] \frac{\rd\bk}{(2\pi)^d} \,  .
\ee
Solving this equation in the Debye approximation, we obtain the critical
temperature
\be
\label{6.11}
 T_\mathrm{c} =  4\pi \, \frac{d-2}{d} \left [
\Gm\left ( 1 + \frac{d}{2} \right ) \right ]^{2/d} J\nu \,  .
\ee
This tells us that $T_\mathrm{c}$ is not defined for $d = 1$ and $T_\mathrm{c} = 0$ for $d = 2$.
In three dimensions, we have
\be
\label{6.12}
 T_\mathrm{c} \simeq 5J \nu \qquad (d = 3) \,  .
\ee

The general equation for the superfluid fraction \cite{Yukalov_14,Yukalov_19}
can be written in the form
\be
\label{6.13}
 n_\mathrm{s} = 1 - \, \frac{Q}{Q_0} \,  ,
\ee
with the classical dissipated heat
\be
\label{6.14}
 Q_0 \equiv \frac{d}{2} \, T \,  ,
\ee
where $d$ is spatial dimensionality, and
\be
\label{6.15}
Q = \frac{\mathrm{var}(\hat P)}{2m N}
\ee
is the actual dissipated heat, expressed through the variance of the momentum
operator
\be
\label{6.16}
\hat\bP \equiv
\int \psi_1^\dgr(\br) (-\ri\vec{\nabla} )\psi_1(\br) \, \rd\br \,  .
\ee
In an equilibrium system, this variance is
\be
\label{6.17}
\mathrm{var}(\hat\bP) = \lgl \hat\bP^2 \rgl \,  .
\ee
Note that the condensed fraction does not contribute to the operator of
momentum (\ref{6.16}) due to the lattice periodicity \cite{Yukalov_19}.

For a three-dimensional cubic lattice, with a lattice spacing $a$, we obtain
\be
\label{6.18}
 Q = \frac{|\bp(\ba)|^2}{2m\rho} \int_{\cB} \,
\frac{\sum_\al \sin^2(k_\al a)}{\sinh^2(\ep_k/2T)} \,
\frac{\rd\bk}{(2\pi)^3} \,  ,
\ee
where the expression
\be
\label{6.19}
|\bp(\ba)|^2 \equiv
\frac{1}{a^2} \, \exp\left ( - \, \frac{a^2}{2l_0^2}\right )
\ee
is used, derived in the tight-binding approximation. Here, the notation
\be
\label{6.20}
l_0 \equiv \frac{1}{\sqrt{m\om_0} } =
\frac{1}{\sqrt{2m}\, (E_\mathrm{R}V_0)^{1/4}}
\ee
means an effective localization length.

For a three-dimensional cubic lattice, the relations
\be
\label{6.21}
a^2 = \frac{3\pi}{2mE_\mathrm{R}} \, ,  \qquad
k_0^2 = 3 \left ( \frac{\pi}{a}\right )^2
\ee
are valid, which yield the ratio
\be
\label{6.22}
 \frac{a^2}{l_0^2} = 3 \pi^2 \, \sqrt{\frac{V_0}{E_\mathrm{R}} } \,  .
\ee
Then, equation (\ref{6.19}) can be written as follows:
\be
\label{6.23}
 |\bp(\ba)|^2 =  \frac{1}{a^2} \exp \left ( -\,
\frac{3\pi^2}{2} \, \sqrt{ \frac{V_0}{E_\mathrm{R}} } \right ) \, .
\ee
Comparing this with the tunneling parameter defined in equations~(\ref{5.3}), we have
\be
\label{6.24}
 |\bp(\ba)|^2 = \frac{1}{2\pi^2} \left (
\frac{J}{aV_0} \right )^2 \,  .
\ee
Therefore, the dissipated heat (\ref{6.18}) is written as follows:
\be
\label{6.25}
 Q = \frac{a}{m\nu} \left ( \frac{J}{2\pi V_0} \right )^2
\int_{\cB}
\frac{\sum_\al \sin^2(k_\al a)}{\sinh^2(\ep_k/2T)} \,
\frac{\rd\bk}{(2\pi)^3} \,   .
\ee

In this way, the self-consistent mean-field approximation allows us to calculate
any thermodynamic characteristic.

\section{Conclusion}

A self-consistent approach, based on the use of a representative
statistical ensemble, developed earlier for uniform Bose-condensed
systems, is extended to Bose atoms in optical lattices. The approach
ensures a gapless spectrum of collective excitations, the
validity of conservation laws, and self-consistent thermodynamics.
It is shown that the approach can be applied to the lattices with
a weak binding as well as with tight binding. For the former case, the
Bloch representation is more appropriate, while for the latter case,
the Wannier representation is more suitable. Both the Bloch and the Wannier representations lead to a similar description. The
results are compared for the self-consistent Hartree-Fock-Bogolubov
approximation. A convenient general formula for the superfluid
fraction of atoms in an optical lattice is derived.

The HFB approximation, used here, is based on the assumption of the condensate
existence, which is taken into account by means of the Bogolubov shift, explicitly
breaking the global gauge symmetry of the system. This approximation, therefore,
is assumed to provide good description, when the Bose condensate is present, and may
be inappropriate when the system passes to an insulating state. This implies that
the HFB approximation for optical lattices can provide an accurate description for
spatial dimensions larger than one ($d > 1$) and nonzero temperatures below the
Bose-Einstein condensation temperature, $0 < T < T_\mathrm{c}$.

The case of zero temperature requires a special consideration. Cubic
optical lattices at zero temperature and unity filling factor $\nu =
1$ have been extensively studied, mainly from the viewpoint of an
insulating state, with the purpose of defining the stability
boundary of this state, corresponding to the critical transition to
the superfluid state. The dimensionless parameter
\[
u \equiv \frac{U}{z_0 J}
\]
has been varied. For a cubic lattice, the number of nearest neighbors is
$z_0 = 2d$. This zero-temperature problem has been treated in the Gutzwiller
approximation \cite{Rokhsar_41,Schroll_42}, dynamical mean-field approximation
\cite{Amico_43}, direct numerical diagonalization \cite{Roth_44}, density-matrix
renormalization group \cite{Kuhner_45}, strong-coupling perturbation theory
\cite{Elstner_46,Damski_47}, and Monte Carlo simulations
\cite{Wessel_48,Capogrosso_49,Capogrosso_50}. The critical values of the above
parameter were found for $d = 1$ as $u_\mathrm{c} = 1.8$, for $d = 2$, as $u_\mathrm{c} = 4.2$,
and for $d = 3$, as $u_\mathrm{c} = 4.9$. The HFB approximation underestimates quantum
fluctuations at zero temperature. That is why it is applicable only for
nonzero temperatures, when thermal fluctuations become more important.

The advantage of using the developed approach for Bose-condensed atoms in optical
lattices at finite temperatures is its relative simplicity, correct gapless
spectrum, the validity of conservation laws, and self-consistent thermodynamics.

\section*{Acknowledgement}

The author is grateful to E.P. Yukalova for useful discussions. Financial support
from the Russian Foundation for Basic Research is acknowledged.


\newpage

\ukrainianpart

\title%
{Самоузгоджений метод для атомів Бозе-конденсату \\в оптичних гратках}
\author[В.І. Юкалов]{В.І. Юкалов}
\address{Лабораторія теоретичної фізики ім. М.М. Боголюбова,
Об'єднаний інститут ядерних досліджень, \\ 141980  Дубна, Росія }

\makeukrtitle

\begin{abstract}
\tolerance=3000%
Розглядаються атоми Бозе в оптичних гратках при низьких температурах
і слабких взаємодіях, коли  конденсат Бозе-Ейнштейна є утворений.
Застосовано самоузгоджений підхід, що базується на використанні
репрезентативного статистичного ансамблю і забезпечує безщілинний
спектр колективних збуджень і чинність законів збереження. Для того,
щоб показати застосовність підходу до обох, слабкого і сильного
зв'язку, проблема розглядається в представленнях Блоха і Ваньє.
Обидва способи приводять до подібних виразів,  що порівнюються з
самоузгодженим наближенням Хартрі-Фока-Боголюбова. Отримано зручну
загальну формулу для надплинної фракції атомів в оптичній гратці.

\keywords конденсат Бозе-Ейштейна, репрезентативний ансамбль,
оптичні гратки, представлення Блоха, представлення Ваньє,
надплинність
\end{abstract}

\lastpage
\end{document}